\documentclass[floatfix,twocolumn,showpacs,prb,amsmath,amssymb]{revtex4-1} 

\usepackage{graphicx} % Include figure files
\usepackage{amsmath}
\usepackage{txfonts}
\usepackage{bm}

\newcommand{\apo}{Ag$_5$Pb$_2$O$_6$}
\newcommand{\sub}[1]{$_{\mathrm {#1}}$}
\newcommand{\subm}[1]{_{\mathrm {#1}}}
\newcommand{\sps}[1]{$^{\mathrm {#1}}$}

\newcommand{\etal}{\textit{et~al.}}

\newcommand{\Tc}{T\subm{c}}

\newcommand{\RH}{R\subm{H}}

\newcommand{\E}{\varepsilon}
\newcommand{\rs}{r\subm{s}}
\newcommand{\eq}[1]{eq.~\eqref{#1}}

\newcommand{\corrected}[1]{#1}

\begin{document}

%\title{Normal-state transport properties of the three-dimensional electron-gas superconductor Ag$_5$Pb$_2$O$_6$}
\title{Transport properties of Ag$_5$Pb$_2$O$_6$: \\
a three-dimensional electron-gas-like system with low-carrier-density}

\author{Shingo~Yonezawa}
\email{yonezawa@scphys.kyoto-u.ac.jp}
\affiliation{Department of Physics, Graduate School of Science, 
Kyoto University, Kyoto 606-8502, Japan}

\author{Yoshiteru~Maeno}
\affiliation{Department of Physics, Graduate School of Science, 
Kyoto University, Kyoto 606-8502, Japan}

\date{\today}

%.Abstract

\begin{abstract}
We report normal-state transport properties of the single-crystalline samples of 
the silver-lead oxide superconductor Ag$_5$Pb$_2$O$_6$,
including the electrical resistivity, magnetoresistance, and Hall coefficient.
From the Hall coefficient measurement, we confirmed that the carrier density of this oxide is 
as low as $5\times 10^{21}$~cm$^{-3}$, one order of magnitude smaller than those for ordinary 
alkali metals and noble metals.
The magnetoresistance behavior is well characterized by the axial symmetry of the Fermi surface and by a single relaxation time.
The $T^2$ term of the resistivity is scaled with the specific heat coefficient, 
based on the recent theory for the electron-electron scattering.
The present results provide evidence that Ag$_5$Pb$_2$O$_6$ is a low-carrier-density three-dimensional electron-gas-like system 
with enhanced electron-electron scatterings.
\end{abstract}

%\pacs{72.15.Qm, 72.80.Ga, 74.70.Dd} 

\maketitle

%%%%%%%%%%%%%%%%%%%%%%%%%%%%%%%%%%%%%%%%%%%%%
%%%%%% Introduction  %%%%%%%%%%%%%%%%%%%%%%%%
%%%%%%%%%%%%%%%%%%%%%%%%%%%%%%%%%%%%%%%%%%%%%

\section{Introduction}

Three-dimensional (3D) interacting electron gas is 
one of the most fundamental subsystems in condensed matter.
Properties of this system are characterized by the carrier density $n$ or by 
the electronic-density parameter $\rs$,
\begin{align}
\rs\equiv \frac{1}{\alpha k\subm{F}a\subm{B}} = \left(\frac{3}{4\pi n}\right)^{1/3}\frac{1}{a\subm{B}},
\label{eq:rs}
\end{align}
where $k\subm{F}=(3\pi^2 n)^{1/3}$ is the Fermi wavenumber, $a\subm{B}=(4\pi\varepsilon_0\hbar^2)/(m\subm{e}e^2)$ is the Bohr radius, and $\alpha \equiv (4/9\pi)^{1/3}$.
Here, $\varepsilon_0$ is the permittivity of the vacuum, $\hbar$ is the Dirac constant, $m\subm{e}$ is the electron mass, and $e$ is the elementary charge.
Crudely speaking, $\rs$ is the radius of a sphere for one conduction electron normalized by $a\subm{B}$.

Interestingly, theories have predicted that the \corrected{electronic} compressibility of the electron gas can be negative for $\rs > 5.25$.~\cite{Takada1991.PhysRevB.43.5962}
Furthermore, the ground state of 3D electron gas can be different from the ordinary Fermi liquid state if $\rs$ is larger:
Ferromagnetic states, insulating states, and even superconducting states have been predicted.~\cite{Takada1978,Rietschel1983,Kuechenhoff1988,Takada1993.PhysRevB.47.5202}
These unconventional phenomena originate from under-screening of the long-range Coulomb interaction.
Recent observation of the negative dielectric constant in low-density Rb fluid demonstrated that the instability of the 3D electron gas against the long-range Coulomb interaction
is not fictitious.~\cite{Matsuda2007,Maebashi2009.JPhysSocJpn.78.053706}

In such a context, superconducting materials with low carrier density are interesting.
Old examples of such low-carrier-density superconductors 
are SrTiO\sub{3-\delta}~\cite{Schooley1964} and Ba\sub{0.6}K\sub{0.4}BiO\sub{3}.~\cite{Mattheiss1988}
Recently, superconductivity in famous wide-gap semiconductors such as diamond,~\cite{Ekimov2004.Nature.428.542} 
silicon,~\cite{Bustarret2006.Nature.444.465} and 
SiC,~\cite{Ren2007.JPhysSocJpn.76.103710,Kriener2008.PhysRevB.78.024517} has been discovered.
Other remarkable examples with a higher critical temperature $\Tc$ are the  
doped $\beta$-HfNCl~\cite{Tou2001} and doped fullerene.~\cite{Hebard1991.Nature.350.600,Tanigaki1991.Nature.352.222}
For these materials, their carrier density $n$ can be as small as the order of $10^{20}$~cm\sps{-3},
which is a few orders of magnitude smaller than $n$ for the typical metals,
e.g. $8.5\times 10^{22}$~cm\sps{-3} for pure copper and $2.5\times 10^{22}$~cm\sps{-3} for pure sodium.~\cite{Ibach-LuethText}

%%%%%%%%%%%%%%%%%%%%%%%%%%%%%%%%%%%%%%%%%%%%%%%%
\begin{figure}[tbh]
\includegraphics[width=8.0cm]{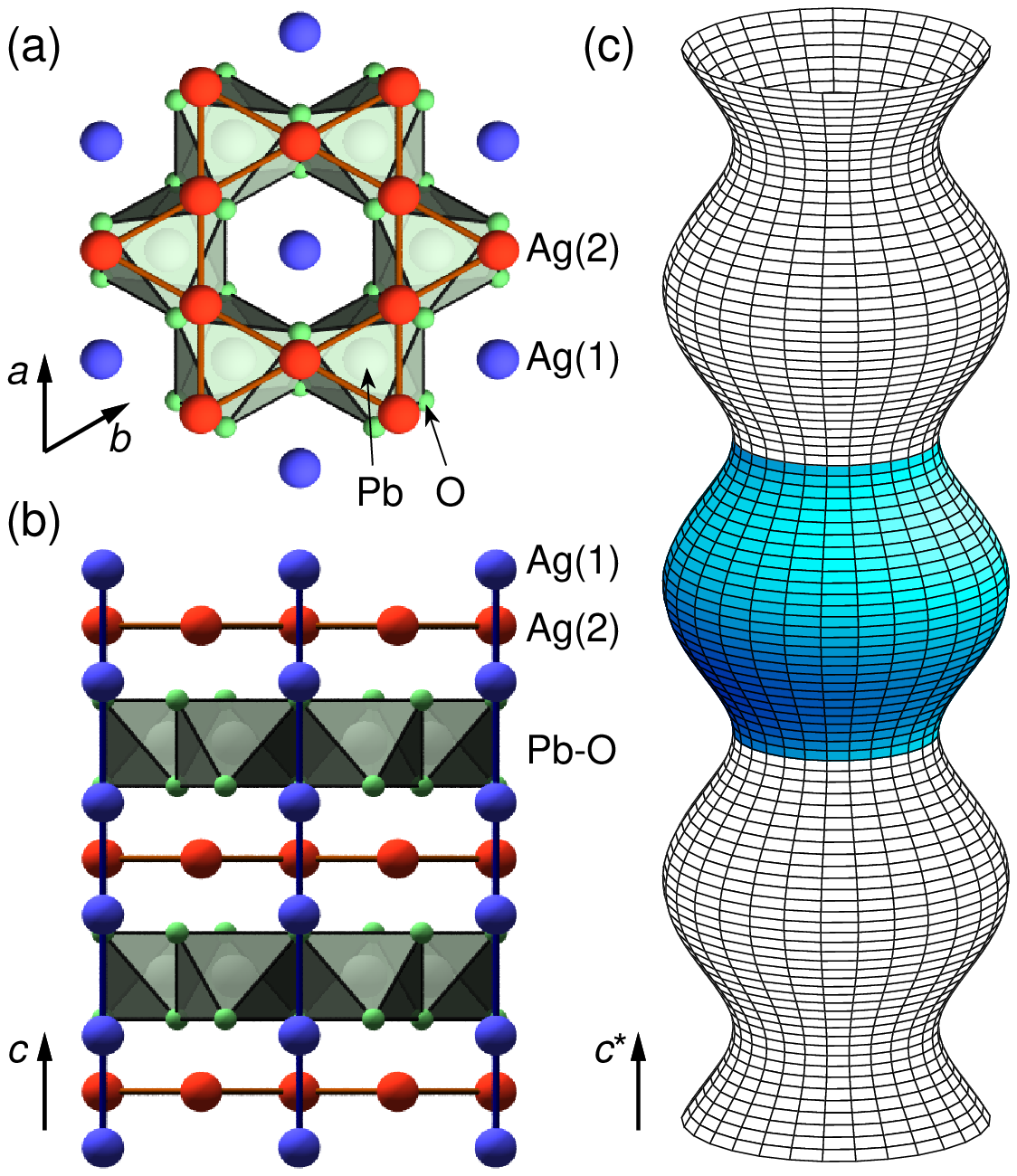}
\caption{(Color online) Crystal structure~\cite{Jansen1990} and Fermi surface~\cite{Oguchi2005,Sutherland2006,Mann2007} of {\apo}.
Panels (a) and (b) present the top and side views of the crystal structure, respectively.
The blue spheres indicate Ag(1) (chain site), the red spheres indicate Ag(2) (Kagome site), and the octahedra indicate the PbO\sub{6} octahedra.
The Fermi surface shown in (c) is drawn using the parameters reported in Ref.~\citenum{Mann2007}. 
The colored region is the Fermi surface in the first Brillouin zone.
\label{fig:Fermi-Surface}}
\end{figure}
%%%%%%%%%%%%%%%%%%%%%%%%%%%%%%%%%%%%%%%%%%%%%%%%

All superconductors listed above are doped semiconductors or insulators.
As a low carrier density superconductor 
with intrinsic carriers (i.e. no carrier doping),
the silver-lead oxide {\apo} ($n\sim 5\times 10^{21}$~cm\sps{-3} and $\rs\sim 6$--8)~\cite{Yonezawa2004,Yonezawa2005} is of a particular importance.
The crystal structure consists of alternating stacking of the Kagome lattice of Ag and honeycomb lattice of PbO\sub{6} penetrated by Ag chains, as shown in Figs.~\ref{fig:Fermi-Surface}(a) and (b).
We discovered the ambient-pressure superconductivity of this oxide below $\Tc=52$~mK
by the resistivity and AC susceptibility measurements.~\cite{Yonezawa2005,Yonezawa2007.PhysicaC.460-462.551}
The band structure calculation~\cite{Oguchi2005} and 
the quantum oscillation experiment~\cite{Sutherland2006,Mann2007}
revealed that {\apo} has a nearly spherical ``free-electron like'' Fermi surface as shown in Fig.~\ref{fig:Fermi-Surface}(c),
despite its apparent layered structure.
\corrected{
The shape of this Fermi surface} is similar to those of the alkali metals and noble metals.
Among such ``simple'' monovalent metals, however, \apo\ is a very rare case of an ambient-pressure superconductor, 
with the only exception of pure lithium with the ambient-pressure $\Tc$ of 0.4~mK.~\cite{Tuoriniemi2007.Nature.440.187}
Moreover, \apo\ is a rare superconductor also in the sense that it is the only-known type-I superconductor among oxides.~\cite{Yonezawa2005}
The band calculation,~\cite{Shein2005.PhysSolidState.47.599,Oguchi2005} as well as 
the neutron diffraction and X-ray absorption studies,~\cite{Yoshii2007}
have pointed out that Pb-6s and O-2p electrons mainly contribute to the electric conductivity; Ag4d and Ag-5s electrons give a minor contribution.
The band calculation suggested that the electronic configuration of this oxide can be represented as Ag\sps{+}\sub{5}Pb\sps{4+}\sub{2}O\sps{2-}\sub{6}e\sps{-}:
i.e. one conduction electron exists per formula unit.
Thus, the electron density $n$ is estimated to be $n\sim 1/V\subm{f} = 5.1\times 10^{21}$~cm\sps{-3}, which is one order of magnitude smaller than $n$ for the alkali metals and noble metals.
Here, $V\subm{f}$ is the volume for one formula unit.

The normal state of {\apo} is also \corrected{very attractive}:
Its resistivity exhibits non-linear $T$ dependence, nearly $T^2$ dependence,
up to room temperature.~\cite{Yonezawa2004}
This dependence is quite different from those of ordinary metals,
whose resistivity depends linearly on $T$ near room temperature
due to the electron-phonon scattering.
The observed $T^2$ dependence suggests a dominance of 
unusual scattering in {\apo}.

In this article, we present normal-state transport properties of {\apo} single crystals,
including the resistivity $\rho$, magnetoresistance (MR) $\rho(H)$, and Hall coefficient $\RH$.
The value of the $T^2$ coefficient of $\rho(T)$ agrees with the recent theory on the electron-electron scattering
by Jacko, Fj{\ae}restad, and Powell.~\cite{Jacko2009.NaturePhys.5.422}
This fact suggests that the origin of the $T^2$ dependence of resistivity is electron-electron scatterings.
The Hall coefficient is negative and its absolute value 
is well consistent with the estimated value of $n$.
The present results experimentally prove that {\apo} is indeed a low-carrier-density analogue of the alkali metals and noble metals, with electron-electron scatterings enhanced by the low carrier density.

%%%%%%%%%%%%%%%%%%%%%%%%%%%%%%%%%%%%%%%%%%%%%
%%%%%% Experiment    %%%%%%%%%%%%%%%%%%%%%%%%
%%%%%%%%%%%%%%%%%%%%%%%%%%%%%%%%%%%%%%%%%%%%%

\section{Experiment}

In the present study, we used single crystals  
of \apo, which have hexagonal rod-like shape along the $c$ direction with dimensions of approximately $1\times 0.2\times 0.2$~mm\sps{3}. 
We grew them by the self-flux method,~\cite{Yonezawa2004}
from mixture of 5-mmol AgNO$_3$ and 1-mmol Pb(NO$_3$)$_2$.
All measurements reported here are performed with a commercial multi-purpose measurement system
(Quantum Design, model PPMS).
The electrical resistivity $\rho$ and the Hall coefficient $\RH$
were measured using a conventional four-probe method with a square-wave current of 
approximately 10~mA,
in a temperature range between 1.8 to 350~K and a field range up to $\pm 70$~kOe.
The electric current was applied parallel to the $c$ axis of the crystal.
To improve the quality of the electrical contacts compared with our previous studies,~\cite{Yonezawa2004,Yonezawa2005}
we used silver paste (Dupont, 4922N) to attach the electrical leads (gold wires) to crystals.
We note that we used diethyl succinate as solvent, 
since some other solvent such as butyl acetate seriously damages the surface of {\apo} crystals.

For Hall coefficient measurements, magnetic field $H$ was applied parallel to the $ab$ plane.
The inter-layer magnetoresistance with $I\parallel c$ was measured in both $H\parallel ab$ and $H\parallel c$.
In the Hall coefficient measurements for $H\parallel ab$, small MR signal is mixed in the Hall voltage $V\subm{H}$,
due to the misalignment of the contact positions.
In order to subtract the MR component from $V\subm{H}$,
we first measured the dependence of Hall voltage on magnetic field from 70~kOe to $-70$~kOe
and then the true Hall voltage $\tilde{V}\subm{H}$ is obtained as 
$\tilde{V}\subm{H}(H)=[V\subm{H}(H)-V\subm{H}(-H)]/2$
based on the assumption that $\tilde{V}\subm{H}$ should be an odd function of $H$.
Similarly, a small Hall voltage in the transverse MR voltage $V\subm{MR}$ for $H\parallel ab$ was subtracted
by extracting the even part of $V\subm{MR}(H)$ on $H$ as $[V\subm{MR}(H)+V\subm{MR}(-H)]/2$.

\section{Results}

\subsection{Resistivity}

%%%%%%%%%%%%%%%%%%%%%%%%%%%%%%%%%%%%%%%%%%%%%%%%
\begin{figure}[tbp]
\includegraphics[width=8.5cm]{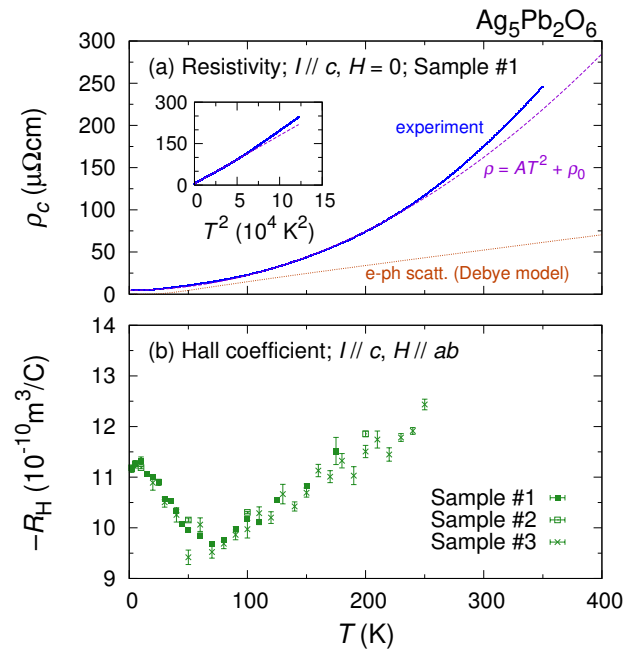}
\caption{(Color online) 
Temperature dependence of (a) the inter-layer resistivity $\rho_c$ and (b) Hall coefficient $-\RH$ of \apo.
The broken curve in (a) is the result of the fitting of the relation $\rho_c(T) = AT^2 + \rho_0$ \corrected{in the range $T < 240$~K. 
The dotted curve represents the resistivity due to the electron-phonon scattering based on the Debye model given by \eq{eq:B-G-formula}. 
In the inset, $\rho_c$ is plotted against $T^2$ with the fitted curve $\rho_c(T) = AT^2 + \rho_0$.}
We note that $-\RH$ is plotted in (b); the value of $\RH$ is negative for the whole temperature range.
\label{fig:resistivity_Hall_Seebeck}}
\end{figure}
%%%%%%%%%%%%%%%%%%%%%%%%%%%%%%%%%%%%%%%%%%%%%%%%

The temperature dependence of the inter-layer resistivity $\rho_c$ in zero field is presented 
in Fig.~\ref{fig:resistivity_Hall_Seebeck}(a).
It is clear that $T$ dependence is non-linear even at room temperature 
and is quite different from
$T$ dependence for ordinary metals: $\rho\propto T^5$ dependence for $T\lesssim 0.2\varTheta\subm{D}$
and $\rho\propto T$ dependence for $T\gtrsim \varTheta\subm{D}$, 
where $\varTheta\subm{D}$ is the Debye temperature.
\corrected{Instead, we can fit the $\rho_c(T)$ data with the formula $\rho_c(T)=AT^2+\rho_0$.
We obtain $A=1.8\times 10^{-3}$~$\muup\Omega$cm/K\sps{2} for the fitting in the range $T<240$~K (see Fig.~\ref{fig:resistivity_Hall_Seebeck}(a)). 
We comment here that a slightly higher value $A=1.9\times 10^{-3}$~$\muup\Omega$cm/K\sps{2} is obtained for the fitting in the whole temperature range (not shown).}

\corrected{
From the fitting, we obtain the residual resistivity $\rho_0 = 4.7$~$\muup\Omega$cm. 
From the relation (c.f. \eq{eq:sigma_H0})
\begin{align}
\rho = \frac{V}{e^2N(\varepsilon\subm{F})\langle v_z^2\rangle\tau},
\end{align}
where $V$ is the unit cell volume, $N(\varepsilon\subm{F})=1.33$~states/eV is the density of states at the Fermi level,~\cite{Oguchi2005} and $\langle v_z^2\rangle = (2.43\times 10^5)^2$~m\sps{2}/s\sps{2} is the squared average of the $z$ component of the Fermi velocity,~\cite{Oguchi2005} we obtain the scattering time at the lowest temperature $\tau = 0.33$~ps
%Furthermore, the low-temperature mean free path along the $c$ axis $l_{c0}$ can be estimated to be $l_{c0} \sim \langle %v_z^2\rangle^{1/2}\tau_0 \sim 39$~nm from the value $\langle v_z^2\rangle^{1/2} = 2.43\times 10^5$~m/s by the band %calculation.~\cite{Oguchi2005}
}

\corrected{
It is important to compare the present results with the Mott-Ioffe-Regel (MIR) limit.~\cite{Gunnarsson2003.RevModPhys.75.1085}
The MIR limit, which is given by $\rho\subm{MIR}\equiv(3\pi\hbar)/(e^2k\subm{F}^2d)$ for a 3D metal, represents the resistivity when the mean free path $l$ becomes comparable with the inter-atomic distance $d$.
Thus, a semiclassical treatment of resistivity, such as the Boltzmann theory, is no longer valid when the resistivity reaches the MIR limit.
Indeed, in many metals, resistivity is found to saturate to a value close to $\rho\subm{MIR}$ at sufficiently high temperatures,
in clear contrast to the prediction of $\rho\propto T$ based on the high-temperature Boltzmann theory.
For \apo, the MIR limit is estimated as $\rho\subm{MIR} \sim 730$~$\muup\Omega$cm, based on the values $k\subm{F}=0.53$~\AA$^{-1}$ determined from the quantum oscillation experiment~\cite{Sutherland2006,Mann2007} and $d\sim a = 5.93$~\AA.~\cite{Jansen1990}
Therefore, $\rho_c$ of \apo\ is still smaller than the MIR limit in the present temperature range, manifesting the validity of semiclassical treatments of transport phenomena.
The absence of any signs of saturation in $\rho_c(T)$ also support this interpretation.
}

\subsection{Hall coefficient}

The magnetic-field dependence of the Hall voltage $\tilde{V}\subm{H}$ was found to be linear for all present temperature and field ranges.
Thus $\RH$ is obtained simply from the slope of the data.
We plot the temperature dependence of $-\RH$ in Fig.~\ref{fig:resistivity_Hall_Seebeck}(b).
The sign of $\RH$ is negative for all temperature range,
indicating that the carriers in this system are electrons.

From the relation $\RH = 1/nq$, 
where $q=-e$ is the charge of the carrier and $e$ is the elementary charge,
we estimate $n$ of {\apo} to be $n=5.7\times 10^{21}$~cm\sps{-3} 
using the low-temperature value $\RH = -11\times 10^{-10}$~m\sps{3}/C.
This value is very close to $1/V\subm{f}=5.1\times 10^{21}$~cm\sps{-3}, 
where $V\subm{f}=195.38$~\AA\sps{3} is 
the volume for one formula unit at 4~K.~\cite{Yoshii2007}
This agreement support the simple picture that
one conduction electron exists in one formula unit.
We note that the band calculation by Oguchi~\cite{Oguchi2005}
predicted the value of $\RH$ to be $-5\times 10^{-10}$~m\sps{3}/C.
This value is nearly half of the observed value,
in reasonable agreement with the present experiment.
\corrected{
We also note that the present value of $n$ provides various superconducting parameters consistent with the observed type-I behavior.~\cite{Yonezawa2005}
}

\subsection{Magnetoresistance}

\begin{figure}[tbp]
\includegraphics[width=8.5cm]{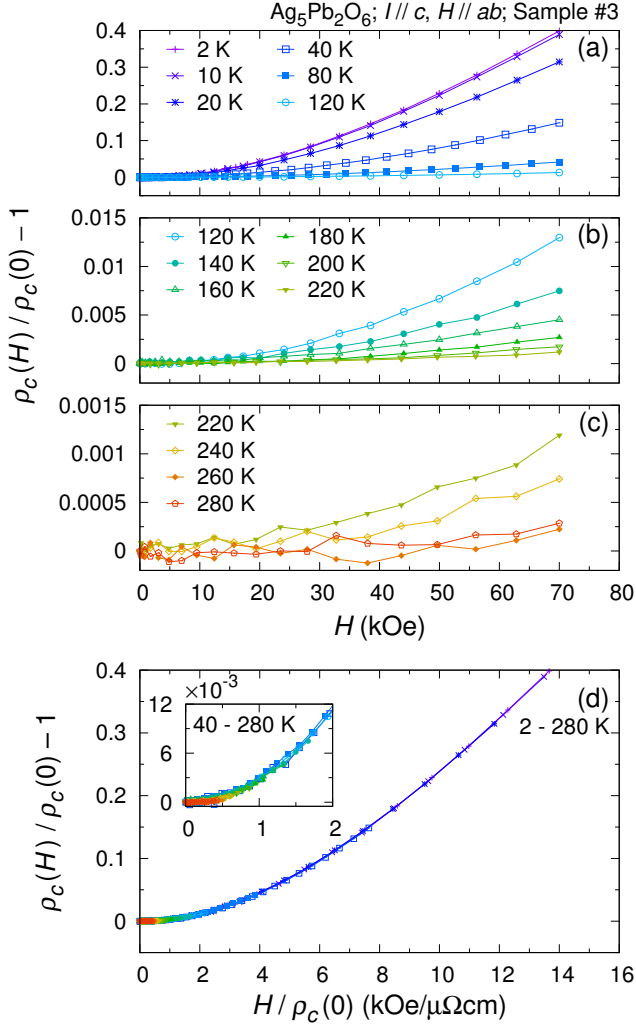}
\caption{(Color online) Inter-plane transverse magnetoresistance (MR) of {\apo} ($I \parallel c$ and $H\parallel ab$). 
Panels (a)--(c) represents the MR data at several temperatures.
The Kohler's plot for all data in the panels (a)--(c) is presented in the panel (d). The inset is the enlarged view near $H/\rho_c(0)=0$ for the data above 40~K.
\label{fig:MR}}
\end{figure}

The magnetic-field dependence of the relative transverse MR $\rho_c(H)/\rho_c(0)-1$ is shown in Figs.~\ref{fig:MR}(a)--(c),
where $\rho_c(0)$ is the resistivity in zero field.
The ratio $\rho_c(H)/\rho_c(0)-1$ is larger for lower temperatures,
and exhibits a quadratic-like field dependence for the present field and temperature range.

It is known that $\rho(H)/\rho(0)-1$ for ordinary metals obeys a scaling law, the Kohler's law,~\cite{ZimanText} with a certain even function $F(x)$:~\cite{Kohler1938.AnnPhys.32.211}
\begin{align}
\frac{\rho(H)}{\rho(0)} -1 = F\left(\frac{H}{\rho(0)}\right).
\end{align}
In Fig.~\ref{fig:MR}(d), $\rho_c(H)/\rho_c(0)-1$ is plotted against $H/\rho_c(0)$.
All the data presented in Figs.~\ref{fig:MR}(a)--(c) collapses into one curve,
indicating that the Kohler's law holds quite well in this oxide.

%%%%%%%%%%%%%%%%%%%%%%%%%%%%%%%%%%%%%%%%%%%%%%%%
%%%%%%%%%%%%%%%%%%%%%%%%%%%%%%%%%%%%%%%%%%%%%%%%

A simple interpretation of the Kohler's law is 
that the scattering process is governed by the product $\omega\subm{c}\tau$,
where $\omega\subm{c}$ is the cyclotron frequency and $\tau$ is the electron relaxation time,
because $\rho(0)\propto\tau^{-1}$ and $H\propto\omega\subm{c}$. 
Thus, the Kohler's law holds when the scattering process is well described with a single $\tau$ over the Fermi surface.~\cite{ZimanText}
The present results indicate that the scattering process remains a single-$\tau$ scattering in the whole temperature range.
%and, possibly, the mechanism of the electron scattering does not vary with temperature.

\begin{figure}[tbp]
\includegraphics[width=8.5cm]{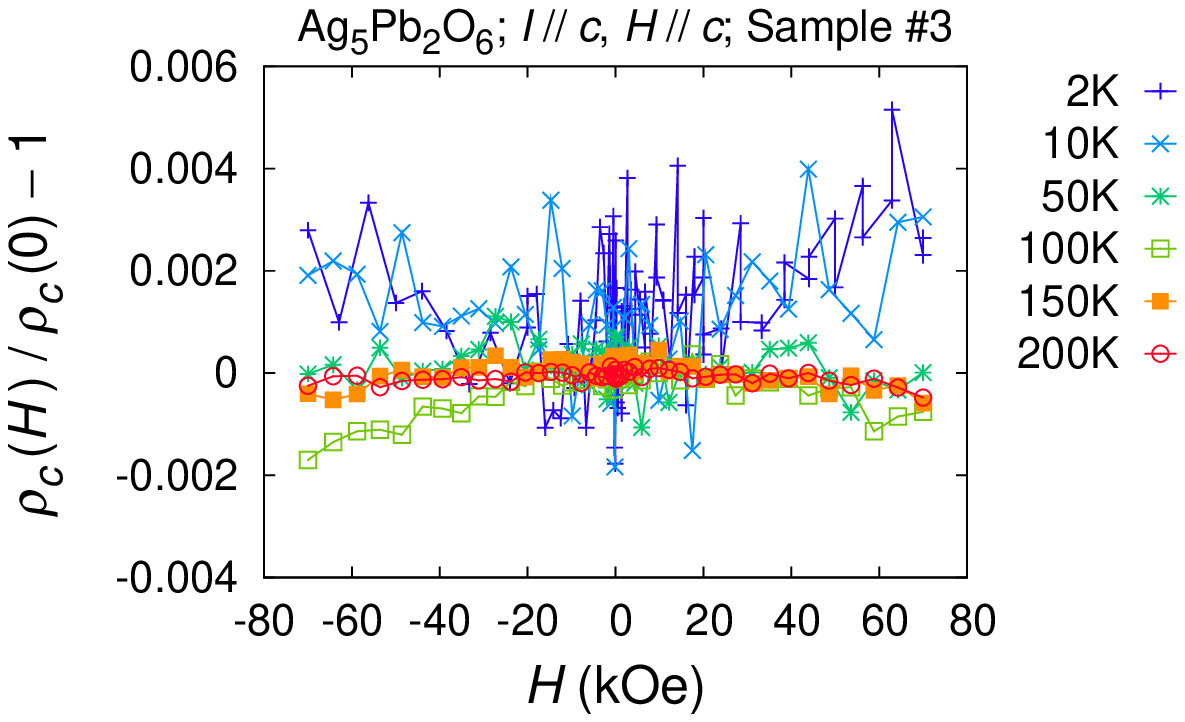}
\caption{(Color online) Inter-layer longitudinal magnetoresistance  (MR) of {\apo} ($I \parallel c$ and $H\parallel c$). 
The apparent negative MR above 100~K is attributed to 
small temperature variation during the measurement due to the MR of the Pt thermometer.
\label{fig:LMR}}
\end{figure}

In contrast, the inter-layer longitudinal MR ($I \parallel c$ and $H\parallel c$) shown in Fig.~\ref{fig:LMR}
is nearly independent of the field strength;
e.g. the longitudinal MR at 2~K is 100 times smaller than the transverse MR in Fig.~\ref{fig:MR}(a).
We also note that the apparent negative MR above 100~K is attributed to small temperature variation during the measurement due to the MR of the thermometer.

%%%%%%%%%%%%%%%%%%%%%%%%%%%%%%%%%%%%%%%%%%%%%%%%
%%%%%%%%%%%%%%%%%%%%%%%%%%%%%%%%%%%%%%%%%%%%%%%%

This absence of the longitudinal MR is well interpreted with the Fermi surface geometry.
Within the semiclassical picture and single-$\tau$ model, the longitudinal conductivity $\sigma_{zz} (\sim \rho_{c}(H\parallel c)^{-1})$ can be expressed as~\cite{Osada1996,Yoshino1999.JPhysSocJpn.68.3027}
\begin{align}
\sigma_{zz} = \frac{\corrected{2}e^2}{V}\sum_{\bm{k}}\left(-\frac{\partial f_0}{\partial\E}\right)
v_z(\bm{k}(0))\int^0_{-\infty}v_z(\bm{k}(t))\exp(t/\tau)\mathrm{d} t,
\label{eq:MR}
\end{align}
where $f_0(\E)$ is the Fermi distribution function and $v_z(\bm{k})$ is the group velocity of the conduction electron along the $c$ axis.
Here, $\bm{k}(t)$ is time dependent due to the cyclotron motion and should follow the equation of motion 
\begin{align}
\hbar\frac{\mathrm{d}\bm{k}(t)}{\mathrm{d}t} = q\bm{v}(t)\times \bm{B},\,\,\,
\bm{v}(t) = \frac{1}{\hbar}\frac{\partial \E(\bm{k})}{\partial \bm{k}}.
\end{align}
For $\bm{B}\parallel c$, the orbital of the cyclotron motion is perpendicular to the $c$ axis.
Because the Fermi surface of \apo\ is nearly axial-symmetric around 
the $c^\ast$ axis~\cite{Oguchi2005,Sutherland2006} (Fig.~\ref{fig:Fermi-Surface}(c)),
the cyclotron orbits are circles around the $c^\ast$ axis in $k$-space and $v_z(\bm{k}(t))$ is time-independent along each orbit.
Thus the right-hand side of eq.~\eqref{eq:MR} reduces to 
\begin{align}
\sigma_{zz} = \frac{\corrected{2}e^2}{V}\sum_{\bm{k}}\left(-\frac{\partial f_0}{\partial\E}\right)
v_z^2(\bm{k}(0))\tau,
\label{eq:sigma_H0}
\end{align}
which is independent of the magnetic field. 
This naturally explains why the field dependence is nearly absent for the longitudinal MR in \apo.

\section{Discussion}

\corrected{
\subsection{Origin of the $\bm{T^2}$ dependence of resistivity \\ - electron-phonon scattering -}
}

From the transport data presented above, we experimentally demonstrate that 
{\apo} is indeed a low-carrier-density 3D electron system with a spherical Fermi surface.
Below, we discuss the origin of the unusual $T^2$ dependence of resistivity.

\corrected{
We first discuss the electron-phonon scattering.
As widely known, the electron-phonon scattering ordinarily results in $\rho\sim T$ at temperatures higher than a few tenth of the Debye temperature $\varTheta\subm{D}$.
In such temperature ranges, the resistivity due to the electron-phonon scattering is given by~\cite{Allen1993,Savrasov1996.PhysRevB.54.16487,Masui2002,NoteCGS}
\begin{align}
\rho\subm{e-p}(T)=\frac{2\pi k\subm{B}T}{\varepsilon_0\hbar\omega\subm{p}^2}
\int^\infty_0 \frac{\mathrm{d}\omega}{\omega}\frac{x^2}{\sinh^2x}\alpha^2F(\omega),
\label{eq:B-G-formula}
\end{align}
where $x=\omega/(2k\subm{b}T)$, $\varepsilon_0$ is the permittivity of vacuum, $\omega\subm{p}$ is the plasma frequency given by $\omega\subm{p}^2 = (ne^2)/(m^\ast\varepsilon_0)$, and $\alpha^2F(\omega)$ is the electron-phonon coupling function.
Here, for the Debye phonon model, $\alpha^2F(\omega)$ can be assumed as $\alpha^2F(\omega) \sim 2\lambda (\omega/\varTheta\subm{D})^4\theta(\varTheta\subm{D}-\omega)$, with $\lambda$ being the dimensionless electron-phonon coupling constant and $\theta(x)$ the step function.~\cite{Allen1993,Masui2002}
By using the values $n=5.7\times 10^{21}$~cm\sps{-3} from the Hall coefficient, $m^\ast = 1.2m\subm{e}$, $\varTheta\subm{D}=186$~K from the specific heat,~\cite{Yonezawa2004}, and $\lambda = 0.29$ from the superconducting transition temperature,~\cite{Yonezawa2004} we obtain the dotted curve in Fig.~\ref{fig:resistivity_Hall_Seebeck}(a), exhibiting a linear temperature dependence with resistivity values much smaller than the observed values. 
Thus, one can conclude that the electron-phonon scattering cannot explain the observed resistivity either qualitative or quantitatively.
As we have discussed in the previous paper,~\cite{Yonezawa2004} a model with higher-frequency optical phonons is not applicable either, since existence of high-frequency phonon modes is totally inconsistent with the specific-heat data.
}

\corrected{Even for ordinary electron-phonon scatterings, effect of thermal expansion may mimic a non-linear temperature dependence of resistivity.
Indeed, such a possibilitiy has been discussed for certain materials such as Rb\sub{3}C\sub{60}.~\cite{Vareka1994.PhysRevLett.72.4121}
However, for \apo, we can show that the thermal expansion effect is not relevant either, as we discuss below.
Generally, the resistivity under constant pressure $\rho_p$ should have a relation with the constant-volume resistivity $\rho_v$ as
\begin{align}
\frac{\mathrm{d}\rho_p}{\mathrm{d}T} = \left(\frac{\partial \rho_v}{\partial T}\right)_p + \left(\frac{\partial \rho_p}{\partial V}\right)_T \left( \frac{\partial V}{\partial T}\right)_p.
\end{align}
We here consider $V$ as the unit-cell volume $V = \sqrt{3}a^2c/2$.
The second term should dominate the first term when a substantial non-linear temperature dependence in $\rho_p$ is observed in spite of a linear temperature dependence in $\rho_v$ expected for electron-phonon scatterings. 
In case of \apo, however, if the second term were comparable to $(\mathrm{d}\rho_p/\mathrm{d}T) \sim 2AT \sim 1.1$~$\muup\Omega$cm/K at 300~K, $(\partial V/\partial T)_p$ should be as large as the order of 0.066~\AA\sps{3}/K from the value $(\partial \rho_p/\partial V)_T \sim 18$~$\muup\Omega$cm/\AA\sps{3} at room temperature deduced from results of hydrostatic-pressure experiments.~\cite{Shibata2002-autumn}
However, this value of $(\partial V/\partial T)_p$ is rather unreasonable, because it is nearly ten times larger than the averaged volume change per one Kelvin $(V\subm{300~K}-V\subm{4K})/(\mathrm{300~K}-\mathrm{4~K})\sim 0.0067$~\AA\sps{3}/K.~\cite{Yoshii2007} 
Nevertheless, we cannot exclude a certain contribution of the thermal expansion in the $\rho_c(T)$ data; for example, a small deviation from the $\rho_c(T)\propto T^2$ behavior above 240~K (Fig.~\ref{fig:resistivity_Hall_Seebeck}(a)) might be attributed to the thermal-expansion effect.
}

\corrected{
\subsection{Origin of the $\bm{T^2}$ dependence of resistivity \\ - electron-electron scattering -}
}

\corrected{Next, we focus on the electron-electron scatterings.}
It has been pointed out by Kadowaki and Woods~\cite{Kadowaki1986} that,
for a class of heavy fermion compounds,
the coefficient $A$ of the $T^2$ term of the resistivity originating from electron-electron scatterings
and the molar electronic specific heat coefficient $\gamma$ 
satisfy a relation $A/\gamma^2=a_0$,
where $a_0=1.0\times 10^{-5}\ \muup\Omega$cm(K\,mol/mJ)$^2$.
The ratio $A/\gamma^2$ is widely known as the Kadowaki-Woods (KW) ratio.
For {\apo}, we demonstrated that the KW ratio is much larger than $a_0$.
From this fact, in the previous paper, we speculated that electron-electron scatterings are not the origin of 
the non-linear temperature dependence of resistivity.~\cite{Yonezawa2004}

One problem of the KW ratio is that it is a ratio between a {\em volume} quantity $A$ and a {\em molar} quantity $\gamma$, as pointed out by Hussey.~\cite{Hussey2005.JPhysSocJpn.74.1107}
Extending this idea, Jacko, Fj{\ae}restad, and Powell (JFP) refined the KW ratio by taking into account the conduction carrier density $n$ and the dimensionality $d$ of the electronic state.~\cite{Jacko2009.NaturePhys.5.422}
They claimed that $A$, $\gamma\subm{v}$, and $n$ hold the relation
\begin{align}
\frac{A}{\gamma\subm{v}^2}\frac{f_d(n)}{4\pi\hbar k\subm{B}^2 e^2} = 81,
\end{align}
where $\gamma\subm{v}$ is the electronic specific heat coefficient for a unit volume.
Hereafter, we shall call the left hand side of this equation as the JFP ratio.
The functional form of $f_d(n)$ varies depending on the dimensionality $d$. 
In the case of $d=3$, this function is
\begin{align}
f_3(n) = \left( \frac{3n^7}{\pi^4\hbar^6} \right)^{\! 1/3}.
\end{align}
Jacko \etal\ uncovered that the JFP relation holds for a wide variety of materials,
including pure metals, transition-metal oxides, heavy fermion compounds, and organic conductors.

\corrected{For \apo, we obtain the JFP ratio of 180--230,
using the values $A=1.8\times 10^{-3}\ \muup\Omega$cm/K$^{2}$, $\gamma\subm{v} = 29.1$~J/K\sps{2}m\sps{3},~\cite{Yonezawa2004} and 
$n=5.1$--$5.7\times 10^{21}$~cm$^{-3}$.}
Although the obtained values are 2-3 times larger than the theoretical prediction,
the value is within the range of the variation of the JFP ratios among different materials, as represented in Table~\ref{tab:JFP}.
We should also note that the coefficient $A$ for the in-plane resistivity $\rho_{ab}$ is approximately two times smaller than that for $\rho_c$.~\cite{Yonezawa2004}
Thus the JFP ratio for $\rho_{ab}$ should be closer to the theoretical value.

%%%%%%%%%%%%%%%%%%%%%%%%%%%%%%%%%%%%%%%%%%%%%%%%
\begin{table}
\begin{center}
\caption{Jacko-Fj{\ae}restad-Powell (JFP) ratios for several three-dimensional conductors compared with the theoretical value. Except for \apo, the values in this table are calculated
from the values listed in the Supplementary Information of  Ref.~\citenum{Jacko2009.NaturePhys.5.422}. 
\label{tab:JFP}}
\begin{tabular}{ccccc}\hline
                    &   $n$ (cm\sps{-3})     &  $A$ ($\muup\Omega$cmK\sps{-2})  & $\gamma\subm{v}$ (JK\sps{-2}m\sps{-3}) & JFP ratio \\ \hline\hline  
\apo                &   $5.1\times 10^{21}$ \footnote{From the chemical formula} &  \corrected{0.0018}                & 29.1                                     & \corrected{180}         \\ 
                    &   $5.7\times 10^{21}$ \footnote{From the Hall coefficient} &  \corrected{0.0018}                & 29.1                                     & \corrected{230}         \\ 
UBe\sub{13}         &   $1.3\times 10^{21}$  &  530                   & 3400                                     & 160         \\ 
LiV\sub{2}O\sub{4}  &   $42.9\times 10^{21}$ &  2                     & 10000                                    & 240         \\
Rb\sub{3}C\sub{60}  &   $4.0\times 10^{21}$  &  0.01                  & 110                                      & 41          \\
Theory& - & - & - & 81 \\ \hline
\end{tabular}
\end{center}
\end{table}
%%%%%%%%%%%%%%%%%%%%%%%%%%%%%%%%%%%%%%%%%%%%%%%%

This agreement with the JFP theory indicates that the electron-electron scattering 
plays a dominant role in the resistivity of \apo.
The disagreement with the KW relation~\cite{Yonezawa2004} is probably due to the low carrier density:
As $n$ decreases, $\gamma$ becomes smaller and $A$ becomes larger, resulting in a KW ratio larger than $a_0$.

Finally, we discuss why the $T^2$ dependence of $\rho$ in \apo\ is observed in the unusually wide temperature range.
Because of the low carrier density in \apo, the electron-electron scattering is enhanced compared to the ordinary metals.
However, as listed in Table~\ref{tab:JFP}, the value of the coefficient $A$ is a few orders of magnitudes smaller than those for typical strongly-correlated systems. 
Thus, the electron-electron scattering in \apo\ is still much {\em weaker} than those in strongly correlated systems.
Nevertheless, in these strongly-correlated systems, $T^2$ dependence is only visible below a certain temperature $T\subm{FL}$,
because strong renormalization of interactions is essential for the $T^2$ behavior.~\cite{Thompson1987.PhysRevB.35.48,Urano2000.PhysRevLett.85.1052}
For example, in heavy fermion systems, $T\subm{FL}$ is closely related to the Kondo effect, which is a renormalization of interactions among the conduction electrons and the localized magnetic moment of $f$ electrons.
In contrast, the electron-electron scattering in \apo\ does not involve such renormalizations.
Therefore, there is no limiting temperature of the $T^2$ dependence in \apo, maintaining the $T^2$ dependence up to above room temperature.

\subsection{Electronic-density parameter $\bm{\rs}$}

We obtain the electronic-density parameter $\rs = 6.5$--6.8 using eq.~\eqref{eq:rs},
from the value $n=5.1$--$5.7\times 10^{21}$~cm$^{-3}$ determined either from the chemical formula or from the Hall coefficient.
As another estimation of $\rs$, we use the relation $\rs = (\alpha k\subm{F}a\subm{B})^{-1}$ and quote the value $k\subm{F} = 0.53$~\AA\sps{-1} determined by quantum oscillations.~\cite{Sutherland2006,Mann2007}
Then, we obtain $\rs = 6.8$.
Furthermore, if we use $a\subm{B}^\ast\equiv (4\pi\varepsilon_0\hbar^2)/(m\subm{e}^\ast e^2)$ with the effective electron mass $m^\ast = 1.2m\subm{e}$ (Ref.~\citenum{Sutherland2006}) instead of $a\subm{B}$,
we obtain $\rs = 8.2$.
We comment here that the observed small mass enhancement is consistent with the Fermi liquid theory for the low-density electron gas
presented by K\"{u}chenhoff and W\"{o}lfle,~\cite{Kuechenhoff1988}
who calculated the mass enhancement $m^\ast/m\subm{e} = 1.02$ for $\rs = 5$ and 1.08 for $\rs=10$.

Theoretically, it is predicted that the electron gas with $\rs > 5.25$ have negative \corrected{electronic} compressibility $\kappa = \partial\mu/\partial n$.~\cite{Takada1991.PhysRevB.43.5962} 
Indeed, experimental evidence for such unconventional electronic state
was reported by Matsuda \etal,~\cite{Matsuda2007} who studied the structure of the low-density Rb fluid in the range $4.5<\rs< 9.0$.
In the case of \apo, $\rs$ is larger than the critical value 5.25. 
So far we haven't observed anomalous behavior indicating negative \corrected{electronic} compressibility.
It is interesting to examine whether the negative \corrected{electronic} compressibility $\kappa$ is realized in \apo. 
The sign of $\kappa$ may be examined from the dependence of the chemical potential on the electron density, e.g. by a quantum-oscillation study under hydrostatic pressure.
In addition, our analysis of the resistivity data indicates strong electron-electron scatterings with light mass dominated by the Coulomb interaction.
Possibility of Coulomb-interaction-induced superconductivity~\cite{Takada1978,Rietschel1983,Kuechenhoff1988,Takada1993.PhysRevB.47.5202} in \apo\ is also worth pursuing.

\section{Summary}

We have studied normal-state transport properties of the 3D electron-gas-like superconductor \apo. 
From the Hall coefficient, we confirmed that the carrier density $n$ is $n\sim 5\times 10^{21}$~cm\sps{-3} and is one order of magnitude smaller than those for alkali metals and noble metals.
The magnetoresistance behavior is well attributed to the approximate axial symmetry of the Fermi surface and to a single relaxation time.
These results experimentally prove that {\apo} is indeed a low-carrier-density analogue of the alkali metals and noble metals.
The $T^2$ coefficient of the resistivity satisfies the Jacko-Fj{\ae}restad-Powell relation.~\cite{Jacko2009.NaturePhys.5.422}
This fact suggest that the origin of the $T^2$ dependence of the resistivity is the electron-electron scattering enhanced by the low value of $n$.

The present value of $n$ corresponds to $r\subm{s}\sim 6.5$--$8.2$.
Thus, crudely speaking, \apo\ is in the regime $\rs>5.25$, where the \corrected{electronic} compressibility can be negative.
It is an interesting future issue to examine the possibility that the negative \corrected{electronic} compressibility is realized in \apo, as well as its relation to the superconductivity.

\begin{acknowledgements}
We would like to acknowledge S.~Nakatsuji and H. Takatsu for their support,
and K. Matsuda, Y. Takada, H. Maebashi, and T. Oguchi for helpful discussions.
This work is supported by the Grant-in-Aid for the Global COE ``The Next Generation of Physics, Spun from Universality and Emergence'' from  Ministry of Education, Culture, Sports, Science and 
Technology (MEXT) of Japan.
It has also been supported by Grants-in-Aids for Scientific Research from MEXT 
and from Japan Society for the Promotion of Science (JSPS).
S.~Y. was financially supported by JPSP.
\end{acknowledgements}

\bibliography{notes.bib,%
D:/cygwin/home/Owner/SSP/paper/string,%
D:/cygwin/home/Owner/SSP/paper/Ag5Pb2O6,%
D:/cygwin/home/Owner/SSP/paper/MgB2,%
D:/cygwin/home/Owner/SSP/paper/oxides,%
D:/cygwin/home/Owner/SSP/paper/TMTSF,%
D:/cygwin/home/Owner/SSP/paper/superconductors,%
D:/cygwin/home/Owner/SSP/paper/Seebeck,%
D:/cygwin/home/Owner/SSP/paper/electron-gas,%
D:/cygwin/home/Owner/SSP/paper/resistivity,%
D:/cygwin/home/Owner/SSP/paper/Type-I_SC,%
D:/cygwin/home/Owner/SSP/paper/textbook}

\begin{thebibliography}{43}
\bibitem [{Not()}]{NoteCGS}%
  \BibitemOpen
  \href@noop {} {}\bibinfo {note} {{Note that resistivity in these articles is
  given in the cgs Gauss unit. Also, the plasma frequency in the cgs Gauss unit
  is expressed as $\omega\subm{p}^2 = (4\pi e^2 n)/m^\ast$.}}\BibitemShut
  {Stop}%
\end{thebibliography}

\end{document}